\documentclass{aastex62}

\def\yohkoh{{\sl Yohkoh}}
\def\hinode{{\sl Hinode}}
\def\sdo{{\sl SDO}}
\def\stereo{{\sl STEREO}}
\def\soho{{\sl SOHO}}

\def\kms{km\,s$^{-1}$}

\def\heii{He~{\sc ii}}

\newcommand{\gtsim}{\raisebox{-1.0ex}{$\stackrel{\textstyle>}{\sim}$}}
\newcommand{\ltsim}{\raisebox{-1.0ex}{$\stackrel{\textstyle<}{\sim}$}}

\long\def\comment#1{}

\graphicspath{{./}{figures/}}

\submitjournal{ApJ}

\shorttitle{Evaluation of the Minifilament-Eruption Scenario for Solar Polar Coronal Jets}
\shortauthors{Baikie et al.}

\usepackage{float}
\usepackage{placeins}

\begin{document}

\title{Further Evidence for the Minifilament-Eruption Scenario for Solar Polar Coronal Jets}

\correspondingauthor{Alphonse Sterling}
\email{alphonse.sterling@nasa.gov}

\author[0000-0002-0786-7307]{Tomi K. Baikie}
\altaffiliation{Previously at: School of Mathematics and Statistics, University of St Andrews, St Andrews,  KY16 9SS, UK.}
 \affiliation{Cavendish Laboratory, J.J. Thomson Avenue, University of Cambridge, Cambridge, CB3 OHE UK.}

\author{Alphonse C. Sterling}
\affiliation{NASA/Marshall Space Flight Center, Huntsville, Alabama 35812, USA}

\author{Ronald L. Moore} 
\affiliation{Center for Space Plasma and Aeronomic Research, University of Alabama in Huntsville, Huntsville, AL 35899, USA}
\affiliation{NASA/Marshall Space Flight Center, Huntsville, Alabama 35812, USA}

\author{Amanda M. Alexander}
\altaffiliation{Previously at: Department of Astrophysical and Planetary Sciences, University of Colorado Boulder, Boulder, CO 80309}
\affiliation{Department of Geological Sciences, University of Colorado Boulder, Boulder, CO 80309}

\author{David A. Falconer}
\affiliation{Center for Space Plasma and Aeronomic Research, University of Alabama in Huntsville, Huntsville, AL 35899, USA}
\affiliation{NASA/Marshall Space Flight Center, Huntsville, Alabama 35812, USA}

\author{Antonia Savcheva}
\affiliation{Harvard-Smithsonian Center for Astrophysics, 60 Garden St., Cambridge, MA 02139, USA}
\affiliation{Bulgarian Academy of Sciences, Institute for Astronomy and National Astronomical Observatory, Sofia,
   Bulgaria}

\author{Sabrina L. Savage}
\affiliation{NASA/Marshall Space Flight Center, Huntsville, Alabama 35812, USA}



\begin{abstract}

We examine a sampling of 23 polar-coronal-hole jets. We first identified the jets in
soft X-ray (SXR) images from the X-ray telescope (XRT) on the \hinode\ spacecraft, over 
2014-2016.  During this period, frequently the polar holes were small or largely 
obscured by foreground coronal haze, often making jets difficult to see.  We selected 23 
jets among those adequately
visible during this period, and examined them further using Solar Dynamics Observatory
(\sdo) Atmospheric Imaging Assembly (AIA) 171, 193, 211, and 304\,\AA\ images.  In SXRs 
we track the lateral drift of the jet
spire relative to the jet base's jet bright point (JBP). In 22 of 23 jets the spire
either moves away from (18 cases) or is stationary relative to (4 cases) the JBP\@.  The one
exception where the spire moved toward the JBP may be a consequence of line-of-sight projection
effects at the limb. From the AIA images, we clearly identify  an erupting minifilament in 20 of the
23 jets, while the remainder are consistent with such an eruption having taken place.  We also
confirm that some jets can trigger onset of nearby ``sympathetic" jets, likely because eruption of
the minifilament field of the first jet removes magnetic constraints on the base-field region of the
second jet.  The propensity for spire drift away from the JBP,  the identification of the erupting
minifilament in the majority of jets, and the magnetic-field topological changes that lead to
sympathetic jets, all support or are consistent with the minifilament-eruption model for jets.

\end{abstract}



\section{Introduction} 
\label{sec:intro}

Solar coronal X-ray jets are features rooted low in the solar atmosphere that extend into the corona.  They have
been observed primarily at soft X-ray (SXR) and extreme ultraviolet (EUV) wavelengths, with the first detailed
studies in SXRs from the Soft X-ray telescope (SXT) on the \yohkoh\ satellite \citep{shibata.et92,shimojo.et96}.
Subsequently, they have been observed with various instruments, including extensively in SXRs with the X-ray
Telescope (XRT) on \hinode, and in EUV with the with SECCHI on the \stereo\ spacecraft and with the 
Atmospheric Imaging Assembly (AIA) on the Solar  Dynamics Observatory (\sdo) satellite, and they have been seen
in images from white light coronagraphs on \stereo\  and \soho\@ \citep[see, e.g.,][]{Raouafi2016}.  The jets
have spires that grow to be long and  narrow; in polar coronal holes they reach lengths of $\sim$50,000\,km with widths of $\sim$8000\,km over lifetimes of $\sim$10-min \citep{savcheva.et07}.  Their  base regions
can become many times brighter than the spires, particularly when observed in SXRs.  Often the base brightening is
asymmetric with respect to the spire, being much brighter on one side of the spire than the other \citep{shibata.et92}; we refer to
the brightest location in the jet base as the jet bright point \citep[JBP;][]{sterling.et15}.  Coronal jets
appear in all regions of  the Sun.  While \yohkoh\ primarily observed jets at the periphery of active regions,
XRT revealed that they are also prevalent in polar coronal holes \citep{cirtain.et07,savcheva.et07}.  They are
also visible in quiet Sun and on-disk coronal holes \citep[e.g.,][]{mcglasson.et19}.  A number of
investigations have studied the on-disk source region of coronal jets, including the magnetic field properties
at the base of the jets
\citep[e.g.,][]{shimojo.et98,huang.et12,shen.et12,shen.et17,young.et14a,adams.et14,panesar.et16a,muglach21}.    See
\citet{shimojo.et00},  \citet{Raouafi2016}, \citet{hinode.et19}, and \citet{shen21} for summaries  and reviews
of coronal jets.

Based on the early \yohkoh\ observations, it was suggested that jets result when closed bipolar 
magnetic field
emerges through the photosphere and into the corona, and undergo magnetic reconnection with an ambient
approximately vertical background coronal field \citep{shibata.et92}.  Numerical simulations based on this
idea reproduce many of the characteristics of jets
\citep[e.g.][]{yokoyama.et95,moreno-insertis.et08,nishizuka.et08}. In this scenario, a bipole  emerging
into a unipolar vertical background field will form a current sheet at an interface between the
majority-polarity background field and the minority-polarity leg of the emerging magnetic arch.  Reconnection at
that current sheet, it was argued, would produce both the JBP and the jet spire.

Following the earliest jet observations, \citet{moore.et10} and \citet{moore.et13} proposed a dichotomy of jets
based on their morphology, primarily when viewed in SXRs.  These observations reveal that the spires of 
some X-ray jets remain thin and ``pencil-like" during their entire lifetime, much narrower than the base of
the jet.  
These were named ``standard jets," because they seemed to follow the basic evolution of the proposed
emerging-flux model, as originally introduced by \citet{shibata.et92}.  

\citet{moore.et10} also however
identified a category of jets that evolve much more explosively and with broad curtain-like spires. 
They proposed that these jets resulted when the field in the core of the emerging (or emerged) bipole
is strongly sheared and twisted relative to its potential-field form, and is triggered to undergo an
eruption that blows out the bipole's entire magnetic arch (hence, ``blowout eruption").  They 
proposed that the blowout eruption is triggered either from inside, at the magnetic neutral line in 
the core of, or from outside of the bipole, by the onset of JBP-making reconnection of
the bipole's magnetic arch with ambient far-reaching field.   Either way, the blowout eruption 
results in a much broader eruption that encompasses the width of the entire jet-base
region, or even broader  than the base.  They called events in this category ``blowout jets," due to their
blowout-explosion nature. Often the exploding bipole would carry cool filament-like material outward as it explodes,
explaining the prevalence of cool EUV-emitting (seen in, e.g., 304\,\AA\ \heii\ images) spires they observed in
these blowout jets. There have been several numerical simulations that reproduce properties of blowout jets
based on the emerging-flux picture \citep[e.g.][]{moreno-insertis.et13,fang.et14,cheung.et15}.

Thus, up until that point, both the standard jets and blowout jets were explained in
terms of the emerging-flux (or recently emerged flux) scenario.  We will discuss the emerging-flux 
model for making jets further in \S\ref{sec-spires}.

Several studies however indicated that, instead, jets are small-scale versions
of solar eruptions  \citep[e.g.][]{nistico.et09,raouafi.et10,hong.et14}, with
some showing clearly small-scale filament eruptions making a jet
\citep{shen.et12}.  Also, some studies found flux cancelation leading to their
observed jets
\citep[e.g.][]{shen.et12,innes.et13,hong.et14,young.et14a,young.et14b,muglach21}. 
Many of these studies utilized high-resolution, high-cadence EUV jet
observations from the Solar  Dynamic Observatory's (\sdo) Atmospheric Imaging
Assembly (AIA), although, even before \sdo's 2010 launch,  there were
indications that jets are made by small-scale flare/CME-type eruptions
\citep[e.g.][]{nistico.et09,raouafi.et10}.

Based on the study of one on-disk jet \citep{adams.et14}, and then of 20 polar-limb jets
\citep{sterling.et15},  \citet{sterling.et15} suggested that most or all coronal jets might be due to
minifilament eruptions,  where the JBP is a miniature version of the flare arcade that accompanies
large-scale filament eruptions, and they presented a schematic for how the minifilament eruptions
produce the jets (cf.~Fig.~2 below). They explained standard jets (i.e., thin-X-ray-spire jets) as
occurring when the erupting minifilament is  largely confined to the base region of the jet; that is,
it is analogous to a scaled-down version of a flare-making confined eruption
\citep[e.g.][]{moore.et01,ji.et03,sterling.et11}, so that little or none of any confined  cool
(mini)filament material leaks out.  Blowout jets would occur when the eruption is  strong enough for
much of the erupting minifilament to escape the base region completely.  Numerical simulations 
by \citet{wyper.et17}
and  \citet{wyper.et18a} successfully modeled the basic scenario for the minifilament-eruption model
for coronal  jets.

In this paper, we follow up on previous studies of observations of multiple coronal jets
near the polar limb \citep{nistico.et09,moore.et10,moore.et13,sterling.et15}, by presenting 
new observations of polar coronal jets using XRT (SXR) and AIA (EUV) images.  We focus on the lateral drift
of the jet spire relative to the JBP (\S\ref{subsec-drift}), which, as we show below (\S\ref{sec-spires}), 
can provide insight into the jet-formation mechanism; 
erupting-minifilament visibility (\S\ref{subsec-minifilaments}); and the
evident sympathetic nature of some jetting events (\S\ref{subsec-sympathy}).

\section{Jet Models and Jet-Spire Drift}
\label{sec-spires}

Movies of coronal jets constructed from XRT images reveal that the jet spires frequently have
transverse drift motions. \citet{shibata.et92} first reported ``translational" motions of jets,
although detailed analysis would have been difficult with the comparatively low-cadence \yohkoh/SXT
data that they used.  With XRT, \citet{savcheva.et07} found transverse drift velocities of
0---35\,\kms; \citet{savcheva.et09} found an average of 10\,\kms\ for the drift velocities, and
further found that in most jets the spire drifted away from the JBP\@.  We present  our analysis
of spire drift in \S\ref{subsec-drift}.  In the present section, we consider what spire drift would be
expected based on jet-production pictures (idealized to 2D\@).  As described above, both the emerging-flux
and the minifilament-eruption models for coronal jets can provide seemingly plausible explanations
for both standard and blowout jets; the question is: how are jets actually made on the Sun?  Here, we
consider the expected spire drift for four cases: (1) standard jets in the emerging-flux model, (2)
blowout jets in the emerging-flux model, (3) standard jets in the minifilament-eruption model, and
(4) blowout jets in the minifilament-eruption model.

\subsection{Spire Drift in the Emerging-Flux Model}
\label{subsec-efr_model}

Figure~1 shows jet formation and the spire drift under the assumption that the spire is directly made by 
reconnection of emerging or emerged closed magnetic field.  Panels~1(a)---1(c) show the jet spire
evolution in a standard  jet.  Panel~1(a) gives the basic setup for the
emerging-flux model.  This picture assumes an ambient background  negative-polarity vertical field, 
representative
of a coronal hole with negative-polarity open field.   For this case, and all the remaining cases, essentially the
same geometry could hold in quiet Sun; the same geometry can hold even for some active-region 
areas, but where the nearly vertical
field might be a far-reaching coronal loop instead of open.  In Figure~1(a), the emerging field forms a
current sheet at its interface with the ambient field, represented by the black dash.  From the continued
emergence (panel~1(b)), {\it external reconnection} (reconnection of the outside of the emerging bipole) 
ensues, producing a small loop (a small arcade in 3D) on the left side.  This reconnection also reconnects
ambient open field from the left side to the right side of the emerging magnetic arch.  The reconnected open
and closed fields are drawn in red.  
Further reconnection (panel~1(c)) grows the small  closed arcade on the left side, and
more reconnected open field that snaps to stand straight up (from magnetic tension) adjacent  to the
previous reconnected open field; this newly reconnected open field stands closer to the loop/arcade than does
the first. \citet{shibata.et92} interpreted the small loop/arcade on the left side  as the JBP observed in
SXRs.  Thus, under these assumptions, the spire drift with time should be toward  the JBP for standard
jets according to the emerging-flux model.

Figures~1(d)---1(h) show a blowout jet based on the emerging flux model.  Here,
it is supposed that the jet  forms when emerging field, or emerged field,
explodes into overlying and surrounding ambient coronal field; this is a
variation of the \citet{shibata.et92} idea introduced  by \citet{moore.et10} and
elaborated upon by \citet{moore.et13}.  \citet{moore.et10} assumed that in this
case the emerging/emerged bipole contains sufficient free energy to power an
explosion of the emerging/emerged bipole.  Panel~1(a) shows the emerging/emerged
field as containing free energy  because of twist in the core field, and this
free energy  is set to be released given the appropriate circumstances. 
Initially reconnection occurs in panel~1(e) at the  external current sheet, as
in the standard-jet case in panel~1(b).  This reconnection is either started by
or triggers the  explosion of the emerging/emerged bipole (panel~1(e)---1(g)). 
The exploding field drives continued external reconnection at the exterior
current sheet, and also drives {\it internal reconnection}, which is
reconnection among the legs of the exploding bipole field (represented by the
lower red X in panels~1(e) and~1(f)).  This driven internal reconnection, along with the
external reconnection, results in a bright base region, where the brightened
region is substantially broader than that from the external reconnection alone. 
Also, the expansion of the exploding bipole results in a broadened spire as it
blows out. Here, the early spire movement is identical to the case of the
standard jet in panels~1(a)---1(c), with the spire marching toward the JBP\@. 
With the continuation of the  explosion of the base bipole, the stacking up of
the field lines progressively closer to the JBP continues.  Hence, just as in
the emerging-flux-model standard-jet case, the spire drifts toward  the JBP for
blowout jets in the emerging-flux model.

Therefore, in the emerging flux (or emerged flux) jet-producing scenario, we
would expect to see  the bright spire drift {\it toward} the JBP with time. 
This result is true for both the so-called standard jet and the so-called
blowout jet cases of the emerging-flux scenario for jet production.

\subsection{Spire Drift in the Minifilament-Eruption Model}
\label{subsec-minifilament-eruption_model}

Figure~2 shows jet formation and the spire drift under the assumption that the
jet is made by a minifilament eruption. Panels~2(a)---2(c) show the spire
evolution in the case where a standard jet forms.  Panel~2(a) shows the 
situation prior to the start of jet formation, where a minority polarity
(positive in this case) flux element resides in a background majority-polarity
(negative) open field region, such as a coronal hole. This minority polarity
might be one end of an emerged bipole or it may have coalesced from smaller
minority-polarity  flux elements \citep{panesar.et18a}, or the minority polarity
may have migrated into the majority-polarity region as a consequence of
photospheric surface flows \citep[e.g.][]{adams.et14}.  As it migrates, it
will reconnect with surrounding negative field, forming an anemone-shaped field
\citep{shibata.et07}. Panel~2(a) shows a cross-section of that anemone field,
where one lobe is more compact due to photospheric converging flows, and has 
sheared magnetic stress build up in it via some process (e.g., photospheric
shearing flows).  Field in the compact side is highly sheared or even
twisted into a flux rope.  Often it contains cool material, forming the
minifilament \citep{sterling.et15}.  At the time of Panel~2(a),  based on our
more recent studies \citep{panesar.et20}, we believe that there is no
substantial current sheet yet present between the minifilament lobe and the
open  ambient coronal field.  So, instead of a dash for a current sheet, here we
represent the location where a magnetic null exists by a dot between the
closed-field lobe and the open field.   There is also a dot at a second null
point (the lower dot), this one interior to the bipole between its legs.  

Panel~2(b) shows the minifilament field erupting, and at this time current
sheets form and reconnections occur at the locations of the dots in panel~2(a): 
internal reconnection  occurring among the extended  legs of the erupting
minifilament field at the location of the lower dot in panel~2(a), and external
reconnection occurring between the outside of the  erupting minifilament-enveloping field and the
encountered ambient field at the location of the upper dot in panel~2(a).  
\citet{sterling.et15}  argue that the lower reconnection product of the 
internal reconnection is responsible for the JBP; it is analogous to the flare
arcade formed below an erupting large-scale filament.   The second (upper)
product of the internal reconnection is closed twisted field added to the
erupting minifilament flux-rope field (this would be a new helical field line in 3D\@). 
The external reconnection also has two reconnection products.  One is a new
closed field line extending over the neighboring  larger lobe (in this 2D
representation) of the anemone region. The second (upper) product of 
the external
reconnection is reconnected open field; the initial reconnected open field  stands (in the corona) close to the location of the JBP\@.  As the minifilament continues to erupt
outward, it moves deeper into the corona in a direction away from the JBP, and
new open field lines stand (in the corona) progressively farther from the JBP\@.  In the
case of the standard jet, the minifilament flux rope does not progress far into the corona
on the far side of the large lobe, and therefore only a narrow spire forms. 
Because in this case the minifilament flux rope does not have enough energy to blowout the enveloping magnetic arcade, it becomes
arrested (confined) in the base of the jet; in this case part of the 
minifilament may fall back to the solar surface and a little of 
it might leak out
into the narrow open spire field.  While the minifilament eruption 
continues, the spire continues to grow, and drifts away from the JBP in this
standard-jet case in the minifilament-eruption model.

Panels~2(d)---2(h) show a blowout jet in the minifilament eruption model.  In this case, the jet starts
(2(d)---2(f)) as in the standard-jet case (panels~2(a)---2(c)), but now the erupting lobe's minifilament flux rope contains enough energy
to expand and to escape  explosively (blow out) from the base region (2(f)), eventually making a wide
splay of field over the base region (2(h)), consistent with the morphology of blowout jets.  As it
becomes larger, the jet spire continues to drift away from the JBP location.  Thus here too, the drift
of the spire is away from the JBP in the blowout-jet case in the  minifilament-eruption model.

Therefore, in the minifilament-eruption jet-producing scenario, we would expect
to see the spire generally drift {\it away} from the JBP, for both the standard-
and blowout-jet cases.  This prediction is in contrast to the flux-emergence
scenario, which predicts  the spire drifts toward the JBP during the jet's
growth.  In 3D, the spire drift could appear differently from the
simplified 2D depiction of Figures~1 and~2; for example, if the spire's drift is
directly toward or directly away from the observer, then drift away from the JBP
would not be apparent.  Usually though, we expect that a polar-coronal-hole-jet
spire that is truly moving away from the JBP to appear to be drifting away 
from the JBP, at least when observed inside the limb, from the perspective of \hinode\ or
\sdo\@.

\section{Instrumentation and Data} 
\label{sec-data}

We will examine jet spire drift and other jet properties in the following sections, using data from
\hinode/XRT and \sdo/AIA\@.  XRT takes images of the Sun with broad-band SXR filters and a pixel size of
$1''.02$ \citep{golub.et07}. Although it is capable of taking full-disk images, for this project we
restricted our data  sets to those with field of view (FOV) smaller than the entire disk and centered on
one or the other of the two polar coronal hole regions.  AIA observes the full solar disk continuously with seven EUV
filters with pixel size of $0''.6$  \citep{lemen.et12}. Because the preferred XRT time periods were
limited to when it was observing polar regions,  we first selected data based on availability of
appropriate XRT data, and then obtained AIA data for matching time periods.

We examined data between 2014 August and 2016 August, and searched for time  periods when XRT was
specifically observing either of the two polar regions.  Almost always this was during the time periods when
\hinode\ Operations Plan (HOP) number 81 was run.  This HOP has run once a month on the two  polar holes
nearly every month since 2008, about two years after \hinode's 2006 launch, and continues through the  time
of this writing.  During these runs,  XRT observes the polar regions roughly continuously for typically six
to eight hours.  There were  breaks in the continuity for short periods due to other pre-planned
observations, and for some periods where \hinode\ experienced spacecraft night \citep{kosugi.et07}.  

Our selected time period of 2014-2016 coincided with a phase of the solar cycle when often the coronal holes
were small and/or often obscured by quiet Sun corona in the foreground.  These effects combine to make it
relatively difficult to observe coronal jets, for example compared to the periods analyzed by
\citet{savcheva.et07}, and \citet{moore.et13} and \citet{sterling.et15}, when the coronal holes were
generally less obscured.  We made an initial selection of 130 candidate jets from over 278 hours of examined
data.  Only  a small percentage of these jets, however, were quite well observed.  We selected from this set
23 jets for the closer analysis presented in this study. Our only criterion was that the jets be well
observed in XRT\@.  While we cannot prove that no bias was involved in our selection, we expect that our
selection is typical of coronal X-ray jets; this is because, based on our visual inspection, our 23 
selected jets appear typical of other well-observed jets that we have seen, including those analyzed in \citet{sterling.et15}. Table~1  presents these jets, along with some of their observed
properties.  Among the remaining 107 jets,  the majority were difficult to examine due to very faint spires,
and/or due the the small (compact) size of the jet.  In the case of the faint spires, often this was likely
a consequence of the haze of the foreground corona that obscured the subtle spire movements that would be
more obvious when the coronal holes are more directly visible.   In the case of the compact sizes, it was
just hard to resolve the structure of the brightening region.  In the majority of cases it would not be
feasible to analyze those jets via the methods we employ here.   Most of our 23 selected
jets were bright enough and prominent enough to show through the coronal haze.   A few of the events, such
as events~12, 15, and 16, occur in coronal holes that are less obscured by coronal haze.  A more
comprehensive study of jets might be attempted specifically focusing on periods of  less-obscured coronal
hole visibility.  

Our observations are with the  Al/poly XRT filter, which is  a relatively ``thin" filter in that it is
sensitive to  SXR emissions from as low as about 1\,MK, which is relatively cool for coronal plasmas, and it
is also sensitive  to all hotter temperatures \citep{narukage.et11}.  This temperature sensitivity is ideal
for  observing jets in polar coronal holes since most of them consist of plasmas in the $\sim$1--2\,MK
range  \citep{pucci.et13,paraschiv.et15}.  The FOV was $384'' \times 384''$, at cadence of 60\,s.

For all of the XRT observing periods, we obtained coincident AIA data with the same (or somewhat  wider)
FOVs. For this project we examined 12-s cadence AIA EUV images at  304, 171, 193, and 211\,\AA, which have a
peak response to plasmas of temperatures $5\times 10^4$, $6.3 \times 10^5$, $1.6\times 10^6$, and $2\times
10^6$\,K, respectively. We selected these  channels, and did not include the ``hotter-response" channels of
131, 335, and 94\,\AA, because \citet{sterling.et15}  found polar coronal hole jets to be best visible in
these cooler channels. In general, the AIA channels can detect mixtures of hotter- ($\gtsim$1\,MK) and cooler-
($< 1$~MK) temperature coronal plasmas due to their broad temperature responses \citep{lemen.et12}.  In
contrast, emissions appearing bright in XRT images unambiguously  come from locations of high ($\gtsim$1\,MK)
temperatures, as it has negligible sensitivity to cooler temperatures.  For this reason,  where there is
ambiguity, XRT images are more appropriate than AIA images for identifying the JBP in most circumstances. 
Given a set of XRT-observed jets with clearly discernible spires,  \citet{moore.et10} and
\citet{moore.et13} showed that only some (roughly $\sim$50\%) jets have clear spires in 304\,\AA\ EUV images,
while others of the set have little or no EUV 304\,\AA\ spire.  A similar detailed  comparison of spire
visibility between SXR jets and other AIA wavelengths has not been carried out, but our experience indicates
that the appearance of jets can vary markedly in different wavelengths and from jet to jet.  Consequently, a
study of AIA-observed spire drift would require separate extensive investigations, and therefore this study
focuses on the spire drift in the XRT images only.


\section{Observations}
\label{sec-observations}

\subsection{Sample Events}
\label{subsec-samples}

From the 23 jets listed in Table~1, we first show two of these jets in detail, one being a standard jet and
the other being a blowout jet.

Figure~3 shows the standard jet, labeled number~1 in Table~1.  Figures~3(a)---3(c) show the jet in SXRs 
from XRT, Figures~3(d)---3(f) show it in EUV from the AIA~211\,\AA\ channel, and Figures~3(g)---3(i) show it in
EUV from the AIA~304\,\AA\ channel.   Since the features are subtle, for these figures (and corresponding videos),
we increase the signal strength  relative to the noise by performing a running sum over two consecutive
images, with the cadence remaining 60\,s for the XRT videos.  For the AIA videos, we present every other
image, so that the cadence is 24\,s, which is fully adequate to show the features while reducing the overall
length of the video to a more manageable duration.  In Figure~3(a), the jet has not started and shows no
signal in the XRT  image. At the same times, there is only a slight brightening in 211\,\AA\ (Fig.~3(d)) and 304\,\AA\
(Fig.~3(g)). By the time of Figure~3(b), the spire is starting to extend outward from the base region.  A
strong brightening, the JBP, appears on the east side of the base. The spire is narrow in width compared  to
the width of the base region (where the base includes the JBP, and the bright loop situated immediately 
west of the JBP; the spire emanates from that west-side bright loop).   In Figure~3(c), the spire has
extended in length,  while retaining its narrow width; this persistent narrow width through the life of the
X-ray jet's spire is characteristic of a ``standard jet," as defined in \citet{moore.et10}.  In 3(c) however, the
spire has drifted westward compared to its location in 3(b); this can be seen by observing the location of
the tip of the spire in the two frames relative to the bright feature at the bottom edge in the middle  of
the panels 3(b) and 3(c). This drifting motion is more apparent in video movie\_fig3 that  accompanies
Figure~3.  This series shows clearly the spire drifting {\it away} from the JBP with time, as is expected in
the minifilament-eruption model.

A minifilament that is starting to erupt is apparent in the AIA 211\,\AA\ (Fig.~3(e)) and 304\,\AA\
(Fig.~3(h)) panels (green arrows).  The images in these panels are a few minutes earlier than that 
in panel 3(b) 
immediately above them, in order to show the spire clearly in Figure~3(b) and the erupting minifilament 
in Figures~3(e) and~3(h).  In AIA 211\,\AA, the video movie\_fig3 shows the spire development and drift, 
in addition to the erupting minifilament.  In AIA 304\,\AA, the video movie\_fig3 shows copious cool 
erupting-minifilament material flowing out from the base region.

Figure~4 shows an example of a blowout jet, this time with images from XRT, and from the AIA 193\,\AA\  and
304\,\AA\ channels (each summed over two consecutive images).   This jet is labeled number~2 in Table~1.  Again in
the early-time images (Figs.~4(a,d,g)), there is little or no indication of an impending jet.  By the time
of Figure~4(b), a strong JBP is apparent, and the jet spire is faint but visibly starting (green arrows). 
By the time of Figure~4(c), the base brightening has expanded beyond the JBP so that the entire base now
is bright, and the spire has expanded to be about as wide as the base.  These characteristics describe 
blowout jets. As indicated by the green arrows in Figures~4(b) and~4(c), the drift of the spire is
largely away from the JBP (black arrow in Fig.~4(b)), which again is consistent with the
minifilament-eruption model.  These motions are more clear in the video movie\_fig4, which accompanies
Figure~4.  Images from the AIA 193\,\AA\ channel  (Figs.~4(e and f), and video movie\_fig4) show
corresponding brightenings and spire drift.

In this case, the erupting minifilament is comparatively faint and diffuse, but it is
visible in the 193\,\AA\ movie, and perhaps more clear in the AIA 304\,\AA\ images (dark 
arrows in Figs.~4(h and i)) and video (movie\_fig4).

\subsection{Spire drift}
\label{subsec-drift}

As argued in \S\ref{subsec-minifilament-eruption_model}, the direction of drift of the spire 
relative to the JBP should reflect the mechanism that produces the jet: the spire should drift
toward the JBP in the emerging-flux model, and it should drift away from the JBP in the
minifilament-eruption model.   For the two events of Table~1 examined in  \S\ref{subsec-samples},
we saw that the spire moved away from the JBP with time.   Here we examine  the spire movement for
all 23 events of Table~1.


We do this by tracking the spire motion relative to the JBP, with assistance of a semi-automated
tracking algorithm in {\it Mathematica}.  We manually selected the JBP and spire to begin
tracking.  We examined the tracking for each timestep, and made manual adjustments to the
automatically selected new spire and JBP locations when necessary (which was frequently the case).

For each jet, we track the spire drift relative to the JBP with this procedure (presented in outline
here; see Appendix~\ref{sec-appenda} for further details): We centered a small blue disk on the JBP, and two small coloured disks on the spire: a red one at the base of the spire and a green one along the  body
of the spire, with a yellow line segment through them that approximately traces the spire.  

Figure~5 shows an example with jet number 8.  The middle panels show the blue disk on the 
JBP, red and green disks on the spire, and the yellow line approximately tracing the spire.  Our
routine then automatically places these markers on the evolved jet in the subsequent frame (which we visually
confirmed  for each frame).  We logged the locations of all the markers, giving us a data base of the spire
drift with time during the jet.  Figure~5's right-most panel shows a plot of the results, with the brown line
showing the distance of the jet-spire base (red disk) from the JBP (blue disk) as a function of time.  When we
initially start tracking, the separation is about $19''$.  There is a rapid drift of the spire away from the
JBP from about t=300\,s to 600\,s, and then the spire drifts much less at later times.  Again the movement is
away from the JBP\@.

Figure~6 shows examples for four more jets, numbers 6, 19, 23, and 9 of Table~1.
The first three cases again show movement of the spire away from the JBP, as in
the case of Figure~5 and the two jets of \S\ref{subsec-samples}.  The
result for Jet~9 (Fig.~6(xiii---xvi)) however is different; Figure~6(xvi) shows that in this case 
the spire moves {\it toward} the JBP (decreasing offset distance with time).  We also see though from Figures~6(xiii---xv) that
this jet is located at the solar limb; in fact, the bottom-most part of the jet base
and the JBP appears to be partially obscured by the limb.  Thus, there is additional uncertainty in 
the drift of Jet~9 in Figures~6(xiii---xvi).



Figure~7 shows the trajectories of the spires relative to the JBPs for the 23 jets
of Table~1.  We have split the results  into two plots for ease of display, where
the jets that we followed for less than 1500\,s are in Figure~7(i), and  the
longer-lived jets in Figure~7(ii).  Red, brown, and purple lines all show cases
where the spire drift was away from  the JBP; we use these three close-but-slightly
different colours for this so the eye  can track which trajectory is which even
when they overlap.  Green tracks show jets for which there was little or no
detectable movement of the spire relative to the JBP, and blue shows any that drift
toward the JBP\@.   This shows that Jet~9 is the only jet in our set that drifts
toward the JBP\@.  All of the  other jets of the plots either drifted away from the
JBP with time (18 jets), or were essentially stationary (mean velocity
$\ltsim$1\,\kms; the four jets marked by the green trajectories) with time.  

Although we cannot confirm this here, we suspect that the motion we detect for Jet~9 is
a consequence of our viewing  perspective -- which places it at or just beyond the solar limb -- rather than
of the spire  actually moving toward the JBP\@.  We believe this projection 
effect to be a plausible explanation since we have observed that
spire motions of on-disk jets can have two components: one component with motion 
directed away from the JBP, and a second component that is normal to that first component. So if we are
observing the over-the-limb jet approximately along the direction of the first component (that is, the
component of the spire's motion is directed roughly toward or away from Earth), then the second
component could make the spire appear to move toward the JBP in some cases, or away from the JBP in other
cases.

Omitting the one negative-drift jet, among the 22 remaining zero or positive-drift jets,  we find
the jet drift speeds (projected against the plane of the sky) to span 0---40\,\kms, with four 
jets that are stationary (which we take to be $\ltsim 1$\,\kms),
and the remaining ones to have an average drift speed of
8\,\kms.  These results are consistent with the the
above-quoted \citet{savcheva.et07} and \citet{savcheva.et09} values of 0---35\,\kms\ for the range 
and average speed of 10\,\kms.

\subsection{Erupting minifilaments} 
\label{subsec-minifilaments}

We inspected the 23 jets of Table~1 for indications of erupting minifilaments.
Jets~1 and~2, discussed in  \S\ref{subsec-samples}, are representative examples
that have  erupting minifilaments visible in AIA images.  Table~1 indicates whether
an erupting minifilament was detected, through visual inspection, near the start of
our jets.  As the notes in Table~1 indicate however, when erupting minifilaments
were observed, they sometimes were better seen in some AIA channels than in
others.  For example, we find the erupting minifilament  for jet~2 to be more
apparent in the 171\,\AA\ and 193\,\AA\ channels than in 304\,\AA, while the
erupting minifilaments for jets 6 and 7 are better seen in 304\,\AA\ than in other
channels.  \citet{sterling.et15} also found erupting minifilaments to be better
seen in 304\,\AA\ than in 171\,\AA, 193\,\AA, or 211\,\AA\ for three of their 25
jets \cite[see the ``minifilament measurement details" section
of][]{sterling.et15}.  Therefore it is necessary to check multiple AIA channels
before concluding definitively that an erupting minifilament may not accompany a
given jet, at least in the case of polar coronal hole jets.   

For at least 20 of 23 of our jets, or 87\%, we could identify an attendant erupting
minifilament.  For two cases, cases 9 and 13, we could not identify an erupting
minifilament.  It was also difficult to see an erupting minifilament in case~16;
although we still suspect that an erupting minifilament was present, it is at best
very faint and  harder to detect than in the other cases.   In both of the two
cases 9 and 13 with no unambiguous erupting minifilament, the jets were at, and
perhaps partially beyond, the limb; thus these jets may have resulted from erupting
minifilaments that were obscured by the limb.  

Our observed correspondence between jets and erupting minifilaments is similar to
that of \citet{mcglasson.et19}, who found evidence for erupting minifilaments
during the early formation time of jets in over 90\% of the 60 on-disk jets that
they examined.   \citet{kumar.et19} report evidence for erupting minifilaments in
67\% of a set of 27 equatorial coronal hole jets that they examined, while they
note that the remaining 33\% showed mini-flare arcades and other eruption
signatures.  While still a majority, their percentage of jets with erupting
minifilaments is substantially smaller than our reported values of in excess of
90\%.  It is unclear, however, whether the \citet{kumar.et19} investigators would
have reported as jets with erupting minifilaments some of the cases where we detect
relatively small erupting minifilaments
(e.g.\ jets 7, 10, 12), or some of our cases where we find erupting minifilaments 
to be relatively indistinct or invisible in some 
of the AIA channels (e.g., jet~16).   Nonetheless, our work here, along with other of
our works \citep[e.g.][]{sterling.et15,panesar.et16a,panesar.et18b,mcglasson.et19}, 
\citet{kumar.et19}, and some earlier works \citep[e.g.,][]{raouafi.et10}, all examined
multiple events and find 
evidence for miniature-flare/CME-type eruptive activity as the cause of nearly all of coronal hole and/or 
quiet Sun jets observed in those studies.

\subsection{Sympathetic Jets} 
\label{subsec-sympathy}

Sympathetic flares are defined as: ``solar flares in different
active regions that apparently occur as the common result
of the activation of a coronal connection between the regions" \cite[see, e.g.,][]{moon.et02}.  There had 
long been suspicion that such sympathetic flare pairs exist,
but proving the case statistically has been a challenge \citep[e.g.][]{pearce.et90,biesecker.et00}. 
Recent observations, however, with  high-resolution and high-cadence full-Sun observations in the EUV
shows that there are many solar eruptions/flares in different active regions that are physically (causally)
connected \citep{schrijver.et11}.

We define ``sympathetic jets" in a fashion similar to the above, as {\it jets that occur  in  different
jet-base regions, apparently as the result of a magnetic connection between  the regions}.  Thus,
sympathetic  jets are when a subsequent jet occurs at a different location due to an apparent magnetic
connection to an earlier jet; we expect that such jets will occur in quick succession.  With the spectral
coverage of XRT alone, it can  be difficult to determine whether two jets in close succession are
sympathetic.  As with the case for sympathetic flares however, upon further investigation with the AIA data,
a causational relationship is sometimes clear.  We found that in our set of jets (Table~1), at least  three
jets may be part of a sympathetic pair of events: 3, 11, and 15.  Jets 3 and 11 each  refer to only one
of the two jets of the pair, as we initially selected only the most obvious of the two in the X-ray images
for closer study.  The AIA data however reinforces that each of these two  jets may have a separate-base
nearby sympathetic counterpart.  For Jet 15, we initially took  both of the sympathetic jets of the pair to
be part of the same single XRT jet.  But upon examination  in AIA images and movies we saw that this event
appears to consist of a connected pair of jets  closely spaced in time but closely offset in base
location.   For at least events 3 and 11, we observe no firm evidence for a physical connection between the
seeming jet pairs.  Event~15 however does seem to consist of two jets from a two-part extended eruption of a
long minifilament. 


We show the jet(s) of event 15 in Figure~8, in XRT images and in AIA 193\,\AA\ and 304\,\AA\ images.  Panels~8(a), 
8(d), and~8(g) show the situation prior to the jets. Panels~8(b) and 8(e) show the first jet, and
8(f) shows the minifilament for that first jet just starting to erupt.  Panels~8(c) and~8(f) show the
second jet, which is displaced by some $40''$ to the west of the first jet's base location, and~8(i)
shows the minifilament that makes that jet just starting to erupt.  Videos in the left, center, and right panels of movie\_fig8 respectively show 
the evolution in XRT, AIA~193\,\AA, and AIA~304\,\AA\@. 

From the 193\,\AA\ video, a microfilament clearly erupts from the first jet's location between
13:21 and 13:28\,UT on 2016 March 17.  Flows from that first-eruption location connect to
a second minifilament, which lifts off and erupts between 13:47 and 13:52\,UT, resulting in the
second jet.

Recently, \citet{tang.et21} report a sympathetic jet pair, consisting of standard-jet and a blowout-jet,
where the standard jet apparently triggered onset of the blowout jet. \citet{chen.et17} reported similar
``complex" jet pairs, whereby a jet occurring on one magnetic neutral line set off a jet on a neighboring
neutral line.  Also, \citet{kumar.et19} report that two of the 27 equatorial coronal hole jets they examined
were sympathetic with eruptions of  (mini)filaments from neighboring bright points.  Moreover, the magnetic
connectivity that we suggest facilitates the sympathetic jets observed here, is consistent with the
suggestion of  \citet{pucci.et12} that similar magnetic connectivity might explain the sympathetic activity
that  they observed among polar jets and multiple X-ray bright points.

\section{Summary and Conclusions} 
\label{sec-conclusions}

From the 23 coronal jets selected from XRT SXR images and examined here, the vast  majority (22 of 23)
showed a spire that drifted away  from the JBP with  time (18 cases) or was virtually stationary (four
cases).   There was only one of the 23 jets where we found the  spire to move toward the JBP with time, and
it was positioned at the solar limb, or maybe even slightly beyond the limb.  Based on the discussions in
\S\ref{sec-spires}, we conclude that the spire drifts support the minifilament-eruption model for solar
jets, where a minifilament eruption makes the jet and where the magnetic field beneath the erupting
minifilament undergoes internal reconnection to produce the (mini)flare-like JBP at or near the
minifilament-eruption-onset location \citep{sterling.et15}.

For the one over-the-limb jet with the spire moving toward the JBP, we have argued that it is
possible that the spire was actually moving away from the JBP with time, but with a large component
of that movement along the line-of-sight of our observations, and  an orthogonal component making the
spire appear to move toward the JBP from our viewing perspective.  Nonetheless, with this data set
alone, we are not able to rule out that this lone case obeys the emerging-flux model, or some 
other mechanism, instead.

\citet{savcheva.et09} studied the motions of polar coronal hole jets using XRT data, and report  that
about 75\%---80\% of the time the spires drift away from the JBP\@.  They do not note whether any in
their sample might have been over the limb.  Also, however, they could not detect spire motions for
about one-half of their sample of over 400 jets, mostly because they were not able to measure motions in
their smaller jets.  Because we report a much lower percentage of stationary spires (four out of 23),
it could be that our detection level of 1\,\kms\ of our 23 selected jets was more sensitive than the
method that they utilized.  Given these considerations, we regard our results for the spire movement
compared to the JBP to be reasonably in agreement with the \citet{savcheva.et09} findings.

We identify minifilaments that erupt at the time of jet onset in 87\% of the 23 jets of
Table~1.  This measurement is also consistent with recent studies of jets
\citep[e.g.,][]{panesar.et16a,panesar.et18b,mcglasson.et19,kumar.et19}.  Thus, this
provides further support for the minifilament-eruption scenario for producing at least a large
majority of coronal hole and quiet Sun coronal jets.  There is also evidence that many 
active region jets also operate via the same processes, although often in a more complex magnetic
environment \citep{sterling.et16b,sterling.et17}.

Our study identifies at least three jets that trigger other jets in nearby, but separate,
low-atmospheric base regions. These sympathetic jet pairs are closely analogous to typical sympathetic
flares.  We have not considered in detail whether it is most appropriate to regard these events as truly
``sympathetic,'' which implies that they occur in photospheric magnetic regions that are largely isolated
from each other, or instead whether ``double eruptions" occurring along different portions of the same
neutral line or on closely adjacent neutral lines, might be a better description.  In the latter case,
situations such as those where multiple eruptions occur in close succession  from the same active region
might be a closer analogy \citep[e.g.,][]{torok.et11,sterling.et04}.  The sympathetic  pair of events
reported by \citet{tang.et21} appear to be such true sympathetic jets, occurring in magnetically separate
locations.  From our inspection of the data presented here, we are left with the impression that such {\it
bone fide} sympathetic jets --- and also such sympathetic jets occurring over substantially larger spatial
scales than our  three examples here --- are very common in the polar regions, but it would be a challenge
to confirm this statistically.

\bigskip

\bigskip

TKB, ACS, and SLS wish to thank Eric Priest in supporting this collaboration, and TKB thanks Akshay Rao for supporting this investigation.  TKB, ACS, and SLS wish to
acknowledge support from the MSFC  \hinode\ Project, and TKB acknowledges additional financial support
through the University of St.\ Andrews.  ACS and RLM were supported by funding from the Heliophysics
Division  of NASA's Science Mission Directorate through the Heliophysics Guest Investigator (HGI) Program.  
AMA acknowledges support from the Research Experience for Undergraduates opportunity funded by NSF Grant No.\
AGS-1460767.   \hinode\ is a Japanese mission developed and launched by ISAS/JAXA, collaborating with NAOJ as
a domestic partner, NASA and STFC (UK) as international partners. Scientific operation of the \hinode\
mission is conducted by the \hinode\ science team organized at ISAS/JAXA\@. This team mainly consists of
scientists from institutes in the partner countries. Support for the post-launch operation is provided by
JAXA and NAOJ (Japan), STFC (U.K.), NASA, ESA, and NSC (Norway).

\appendix 

\section{Semi-Automatised Tracking Algorithm} 
\label{sec-appenda}

Our semi-automatised tracking algorithm tracks a jet spire's lateral offset from  the JBP in XRT images.
The user inputs the location of the JBP and spire at a specific time, typically near the maximal brightness
of the jet. In a blowout jet in which the spire has multiple strands at its maximum brightness in XRT images,
we input the location of the spire's brightest strand.  (The only exception to this was event 16, for which 
the two sides of the spire were about the same intensity; in that case we opted to  track the poleward side
of the spire, but selecting the side opposite the pole would have  produced similar results, because motion
of the edges of the spire are approximately  symmetric with respect to the JBP for that jet.) The algorithm
binarises the XRT imagery, defining regions by a global intensity threshold. This image is then passed
through a filter deleting small components, typically on the order of 3-5 pixels. The resulting image
typically results in a approximately circular region corresponding to the jet bright point and a thin
rectangular section defining the jet spire. The algorithm tracks the circular region corresponding to the JBP
and the thin rectangular region corresponding to the jet spire. It does so by monitoring the set of pixels
surrounding the spire and JBP at one time step, and looks for connected sets of pixels that undergo an
intensity increase in the new time step. These typically form well defined  shapes, which assists in the
feature identification for the subsequent steps.  The algorithm monitors the base of the spire, and the
centroid of the JBP defined by its bounding radius, for times before and after the specific time initially
chosen by the user. The algorithm then places coloured markers, as in figures~5 and~6, for user evaluation.
Due to fluctuations in the imagery, and the weak contrast, particularly at the end and beginning of the jet
event, user input is needed in most cases to maintain acceptable accuracy in the trace. The algorithm
potentially may be improved using machine learning, where we may begin to be able to identify images directly
without the binarisation process and small-component filter treatment.

\bibliography{ms_arxiv_220121}

\clearpage

\newpage
\comment{

Omitted stuff.  For details, see 210125 and earlier versions.
} 

\newpage

\startlongtable
\begin{deluxetable}{llllll}
\tablewidth{290pt}
\tabletypesize{\scriptsize}

\tablecaption{Jets selected for further analysis \label{table4}}

\tablehead{
 \colhead{XRT Event} & \colhead{Date\tablenotemark{a}}  & \colhead{Time\tablenotemark{a}} &\colhead{Coordinates\tablenotemark{a}} &\colhead{Erupting Minifilament (EMF) Visibility}   \\
}
\startdata
1 & 2014 August 13& 09:41&  (-40, -760)  & Clear EMF in all channels. \\
2 & 2015 February 27& 12:37&  (10, -840) & EMF, best in 171 and 193\,\AA\@. \\
3 & 2015 March 14& 13:18&  (-110, -780) & Clear EMF in all channels, probably confined; rolling motion.\\
4 & 2015 March 14& 14:15&  (-10, -860) & EMF, best in 193\,\AA; probably confined at low height.\\
5 &  2015 March 17& 13:30&  (-170, -760) & Clear EMF in all channels.\\
6 & 2015 March 19& 12:51&  (-160, -820) & EMF in all channels, best in 304\,\AA\@.\\
7 & 2015 March 23& 17:27&  (180, -870) & Small EMF, best in 304\,\AA\@. \\
8 & 2015 April 8& 13:50& (-100, -810) & EMF in all channels. \\
9 & 2015 June 5& 13:06& (180, -870) & No unambiguous EMF detected.\\
10 &  2016 January 7& 02:25 & (-60, -970) & Small EMF, maybe best in 211\,\AA\@.\\
11 & 2016 January 7& 05:04& (-130, -820) & Clear EMF, likely confined.\\
12 & 2016 March 8& 12:22&  (140, -930) & Small EMF, maybe best in 171 and 193\,\AA\@.\\
13 & 2016 March 8& 15:15&  (20, -970)  & No EMF visible (perhaps over the limb).\\
14 & 2016 March 14& 16:06&  (30, -830)  & Clear EMF, best in 171 and 304\,\AA; maybe confined.\\
15 & 2016 March 17& 13:53&  (80, -880) & EMF, clear in all channels.\\
16 & 2016 March 17& 16:06& (50, -890) & Suspected EMF, faintly visible in 211\,\AA\ only (over $\sim$15:45---15:55\,UT\@). \\
17 &  2016 March 23& 14:09&  (70, -920) & EMF; small size in some channels, larger in 304\,\AA\@. \\
18  &  2016 March 23& 14:33& (-30, -950) & EMF, probably best in 211\,\AA\@.\\
19  &  2016 March 23& 15:19& (70, -920) &  EMF, probably confined.  Clear in all except 304\,\AA\@.\\
20  & 2016 April 01& 10:30&  (-130, -930) & EMF; smallish, visible in all channels, maybe confined.\\
21  & 2016 April 01& 11:02&  (-170, -830) & EMF in all channels, probably partially confined to low height.\\
22 &2016 April 01& 11:36&  (-170, -890) & EMF, faintly visible in all channels. \\
23 &2016 April 01& 11:55&  (-170, -830) & EMF, clear in all channels. Partially confined.\tablenotemark{b}\\
\enddata

\tablenotetext{a}{Determined from XRT images, for a time when the jet is well developed.}
\tablenotetext{b}{EMF is initially confined, but cool minifilament material may leak into spire near the end of the jet.  This jet is 
nearly homologous with 21, but the minifilamement eruption originates from a slightly different location.}
\end{deluxetable}
\clearpage

\begin{figure}[] 
\includegraphics[width=0.9\textwidth] {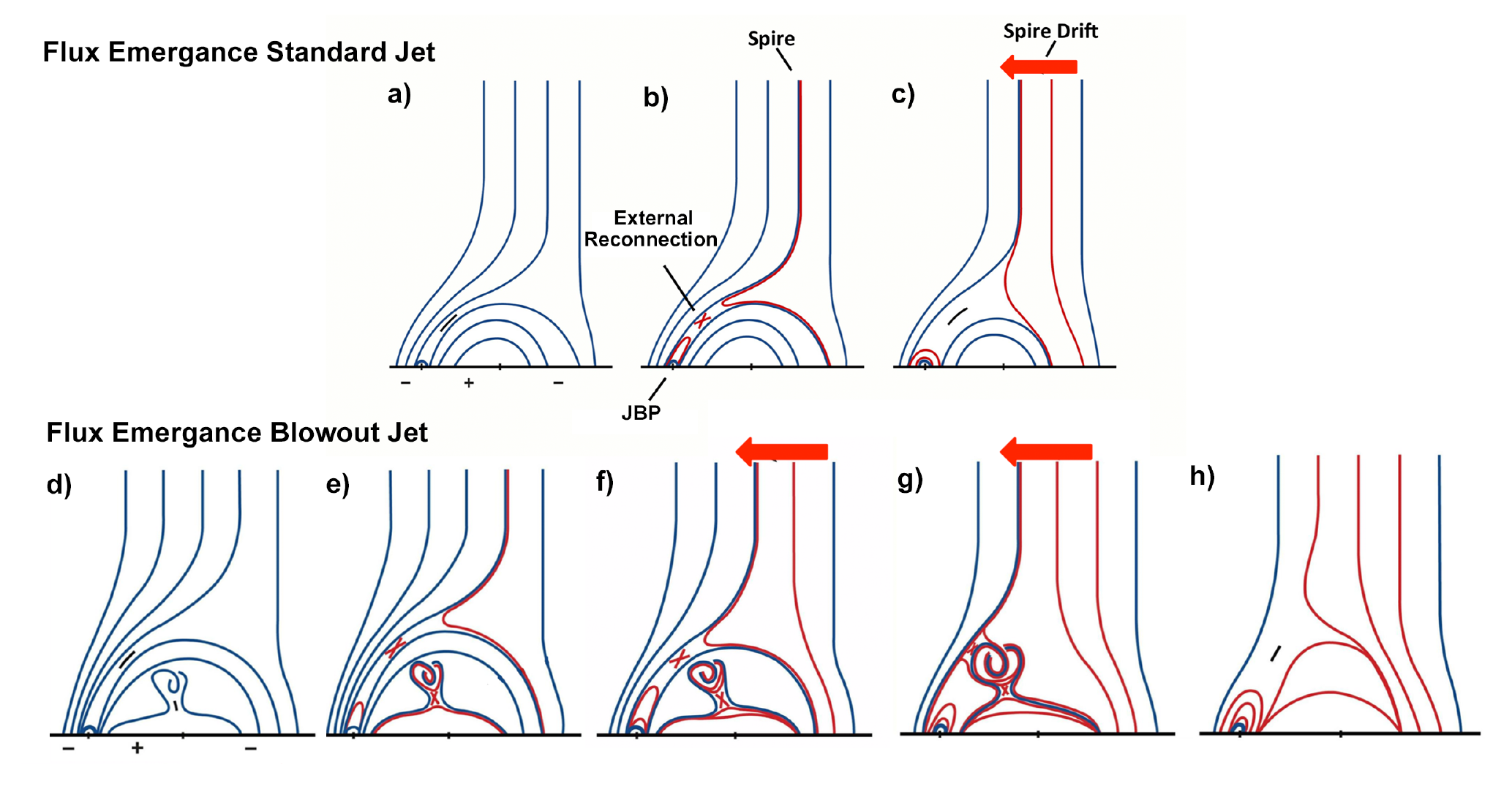} 
\caption{A cartoon depicting the emerging-flux model, as discussed in \S\ref{subsec-efr_model}, with a new bipole
emerging into an ambient open or far-reaching coronal field. Blue lines are not-yet-reconnected field lines 
and red lines are reconnected field lines.  Panels (a)---(c) show development of a standard jet, where reconnection at the
current sheet (short dash in (a); also in (d)) results in the jet bright point (JBP) and 
reconnected open field along the spire. 
The reconnected open field lines stack up closer and closer to the JBP as the reconnection continues, 
resulting in the spire drifting towards the JBP\@.  Thus, the lateral
distance between the JBP and the spire decreases with time, as indicated by the red arrow.  
Panels
(d)---(h) show formation of a blowout jet according to the emerging-flux model.  This time the emerging
field contains non-potential free energy, represented by a twisted field inside of the emerging bipole. 
Again the spire would be expected to migrate toward the JBP, as indicated by the red arrows.  This picture
was introduced and further developed in \citet{shibata.et92} and \citet{yokoyama.et95}, and modified by 
\citet{moore.et10}.  Original versions of these
cartoons are from \citet{moore.et15}.} 
\label{jet_movement_1} 
\centering 
\end{figure} 
\clearpage

\begin{figure}[]
\includegraphics[angle=90,width=0.95\textwidth] {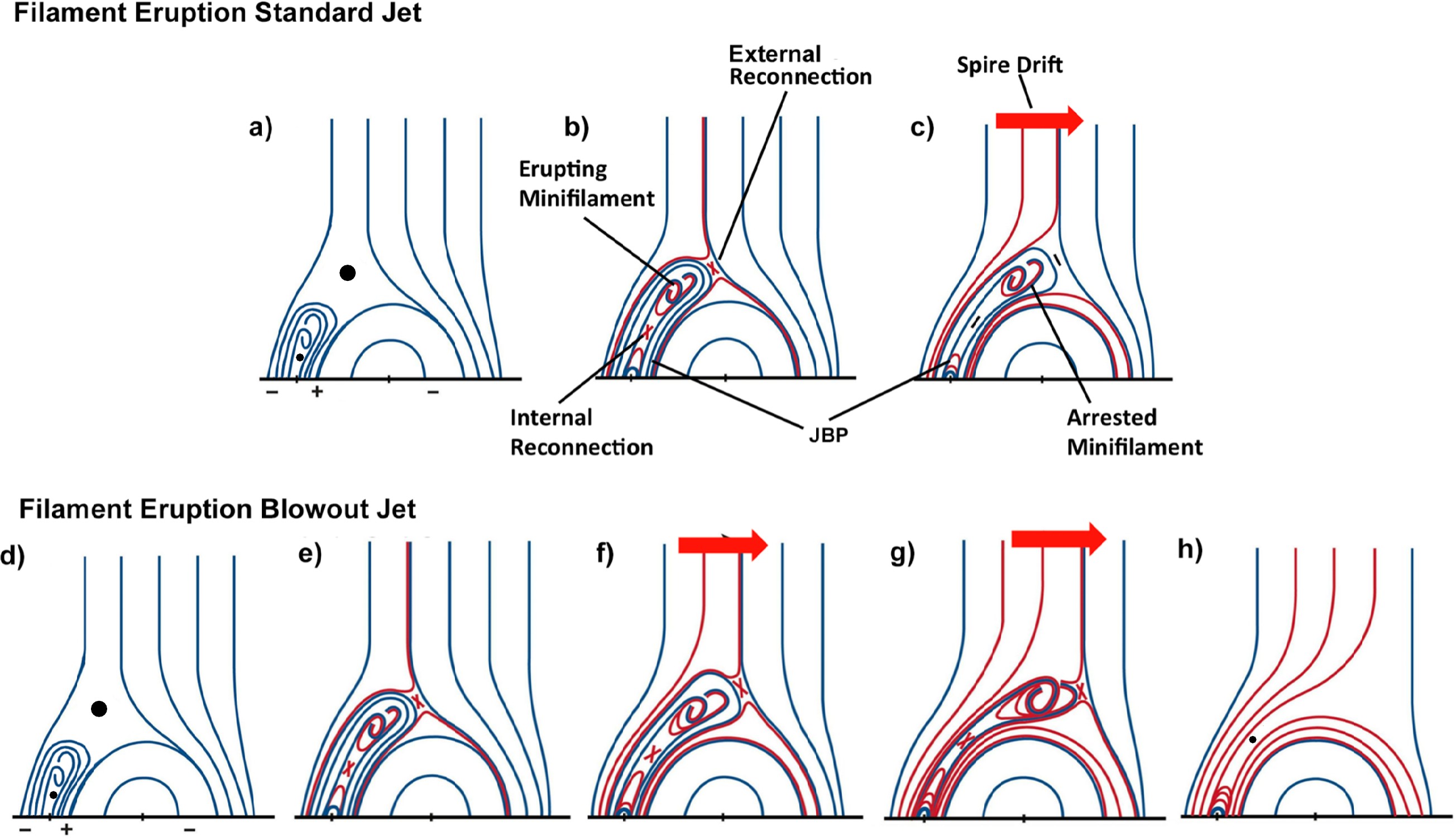}
\caption{A cartoon depicting the minifilament-eruption model, as discussed in
\S\ref{subsec-minifilament-eruption_model}.  Blue and red lines represent magnetic fields as
described in Fig.~\ref{jet_movement_1}.  This time, a minifilament-carrying  bipolar field
erupts from the left side of the base region. Dots in (a), (d), and (h) represent locations 
of magnetic nulls, as our expectation is that substantial current sheets are not present at
these locations at these times.    Panels (a)---(c) show the situation for a
standard jet, where the eruption is largely arrested, resulting in a narrow spire.  Panels
(d)---(h) show the situation for a blowout jet, where the minifilament eruption is strong
enough for the minifilament-flux-rope to be completely consumed (opened) by reconnection with the
ambient  field.  In this case the erupting field moves deeper into the open field as the
filament erupts outward.  In this model, the JBP is a miniature flare arcade that results 
from internal reconnection 
at the location of the lower dot in (a) and (d).
 In both of these  cases (standard jet and blowout jet), the spire
moves away from the JBP as the jet develops. This picture was developed by
\citet{sterling.et15}.  Original versions of these cartoons are from \citet{moore.et15}.}
\label{jet_movement_2}
\centering
\end{figure}
\clearpage

\begin{figure}
\hspace*{2.2cm}\includegraphics[angle=0,scale=0.80]{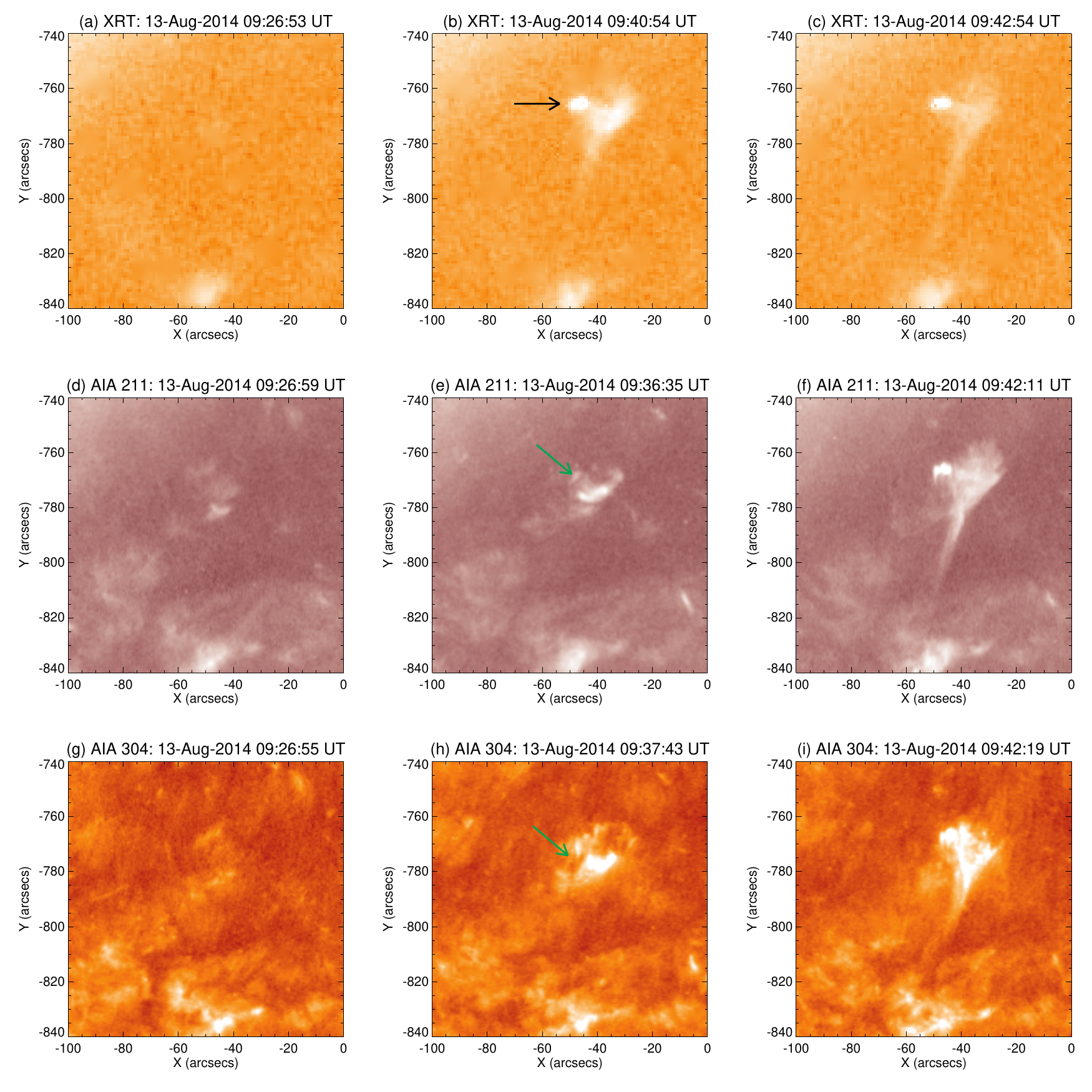}\vspace{1.0cm}
\caption{
Example of a standard jet, Jet~1 in Table~1, as observed in \hinode/XRT Al/poly (a---c),
and \sdo/AIA~211\,\AA\ (d---f), and 304\,\AA\ (g---i) images.   The XRT and AIA images are formed
by summing two consecutive images to increase the signal strength. In this case, in the XRT images 
the spire remains narrow and the brightest part of the base is the
JBP, indicated by the black arrow in (b).  Green arrows in (e) and
(f) show a dark minifilament that starts to erupt from the location of the JBP
in (b).  Times of the vertical column of images in (a), (d), and (g) are the same within a few seconds, and
the same is true for (c), (f), and (i).  Panels (e) and (h) are at about the same time,
but a few minutes prior to (b); this is so that (b) can highlight the JBP and (e) and (h)
can highlight the erupting minifilament.  North is upward and west is to the right in this and in 
all solar images of this paper.  Accompanying videos are in movie\_fig3. These videos are constructed by performing a running sum of every
two consecutive images.  For XRT we show the full-cadence (60\,s) running-sum movie.  For AIA we
show every-other frame of the running sum, resulting in a cadence of 24\,s, which is still adequate 
to show details of the jet evolution while keeping the duration of the movie to a more 
convenient length.}

\end{figure}
\clearpage

\begin{figure}
\hspace*{2.2cm}\includegraphics[angle=0,scale=0.80]{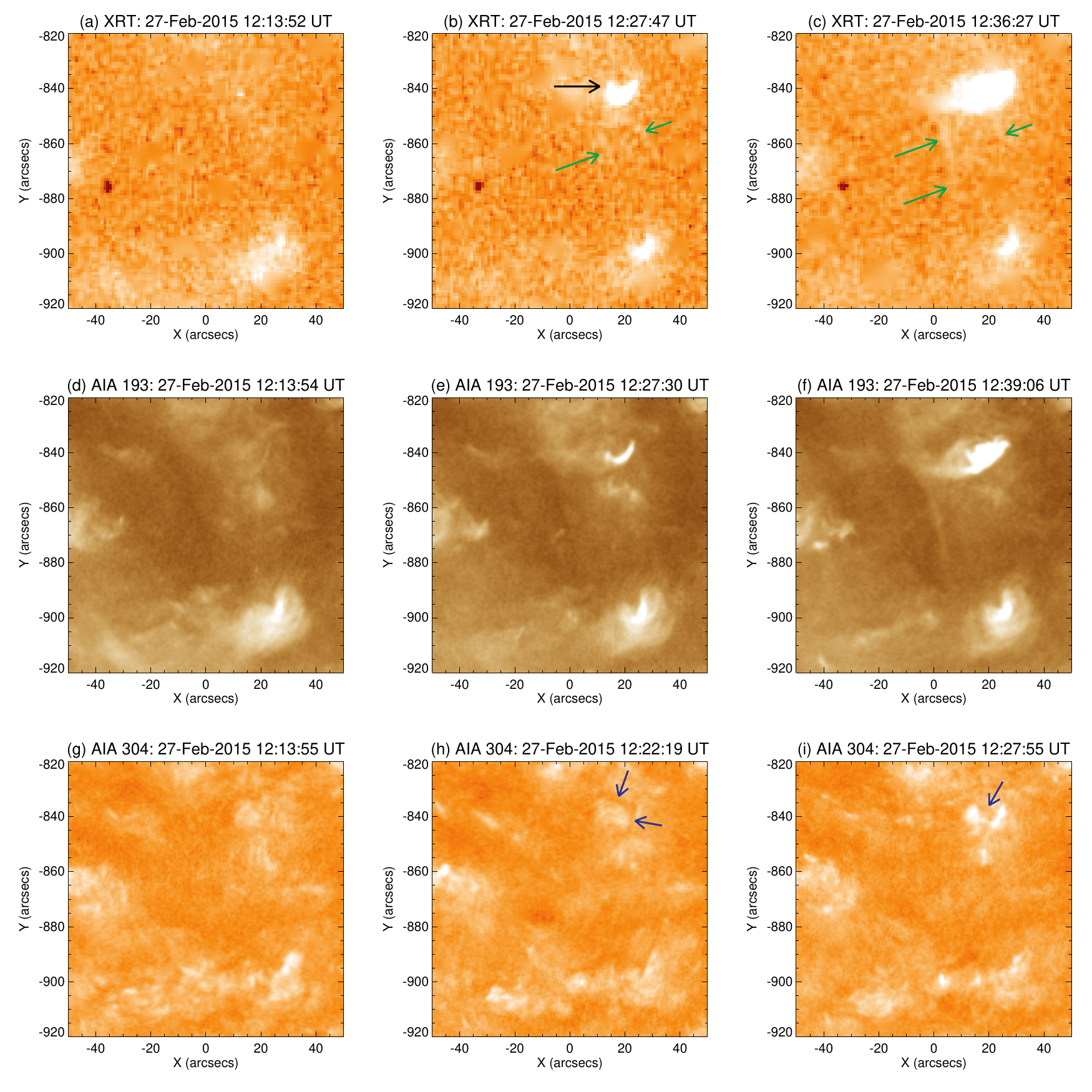}\vspace{1.0cm}
\caption{
Similar to Figure~3, but for a blowout jet, Jet~2 in Table~1, as observed 
in XRT Al/poly (a---c), AIA~193\,\AA\ (d---f), and AIA 304\,\AA\ (g---i) images.   The 
XRT and AIA images are formed by summing two consecutive images to increase the signal 
strength. The black arrow in (b) shows the JBP, and green arrows in (b) and (c)
show the jet spire.  In this case, the entire base brightens between 
(b) and (c), and the spire broadens and eventually becomes about as wide as the 
base, as in panel~(c).  AIA 193\,\AA\ images (d---f) show similar features.  A faint erupting 
minifilament is visible in AIA 304\,\AA\ (dark blue arrows in h and i); it is also visible 
in 193\,\AA\ in the accompanying video (movie\_fig4).  Panels in the same vertical column are
close in time, with some panels in the same column offset by a few minutes from others
in order to highlight respective features.  Accompanying videos are in movie\_fig4.  These are running-sum movies, displayed as described for
the Fig.~3 movies with cadences of 60\,s and 24\,s for XRT and AIA respectively.   In this 
figure, as well as in Fig.~8, we use a non-standard colour table for
the AIA 304\,\AA\ images (and corresponding videos), as we found our selected colours to show the 
erupting minifilaments more clearly for these cases than when rendered with the standard 
colour table.}

\end{figure}
\clearpage

\begin{figure}[H]
\includegraphics[width=1.0\textwidth] {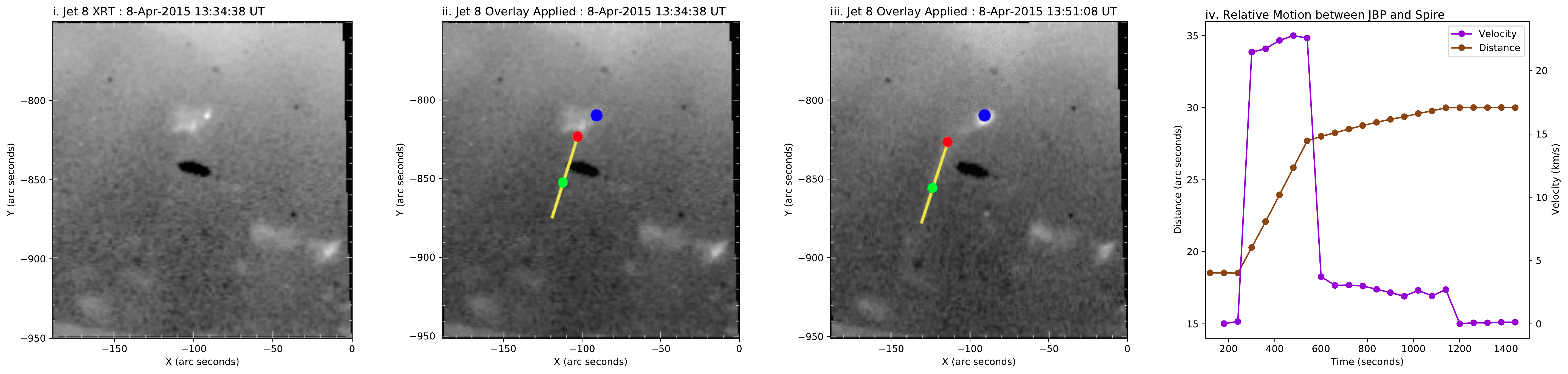}
\caption{\hinode/XRT imagery of Jet 8 of Table~1. The left three  panels show XRT images of the jet, 
with the left-most panel showing the image alone.  In the next two panels, the location selections 
for the JBP (small blue disk), jet-spire base (small red disk), and a point along 
the spire (small green disk) from the jet-spire-tracking routine are overlaid, along with a yellow line 
approximately 
overlying the spire.  Dark spots, including the black oval near the spire, are ``spot" artifacts; in normal
processing these are largely removed (e.g.\ XRT panels in Figs.~3, 4, and 8), but those corrections were 
avoided for the tracking so that real-but-subtle features would not be removed by the despotting 
process. In the right-most panel, the brown line gives the 
resulting track of the spire base point (red disk in the second and third panels) relative to 
the JBP location (small blue disk in the second and third panels).  The purple line
shows the speed of the spire base point relative to the JBP in \kms, determined by effectively taking the derivative of values of the brown line. The start time is 13:31:31~UT on the date of the jet. The shape of the velocity curve is characteristic of many that we observed:
first there is an initial slow drift (until time $\sim$300\,s), and then a sudden fast drift 
($\sim$300---600\,s), 
and then a slowing to a low drift speed ($\sim$600---1100\,s), and finally a slowing to near-zero 
drift speed ($\sim$1200---1400\,s).}
\label{al_215_panel}
\centering
\end{figure}
\clearpage

\begin{figure}[H]
\hspace*{0.0cm}\includegraphics[angle=0,width=1.0\textwidth] {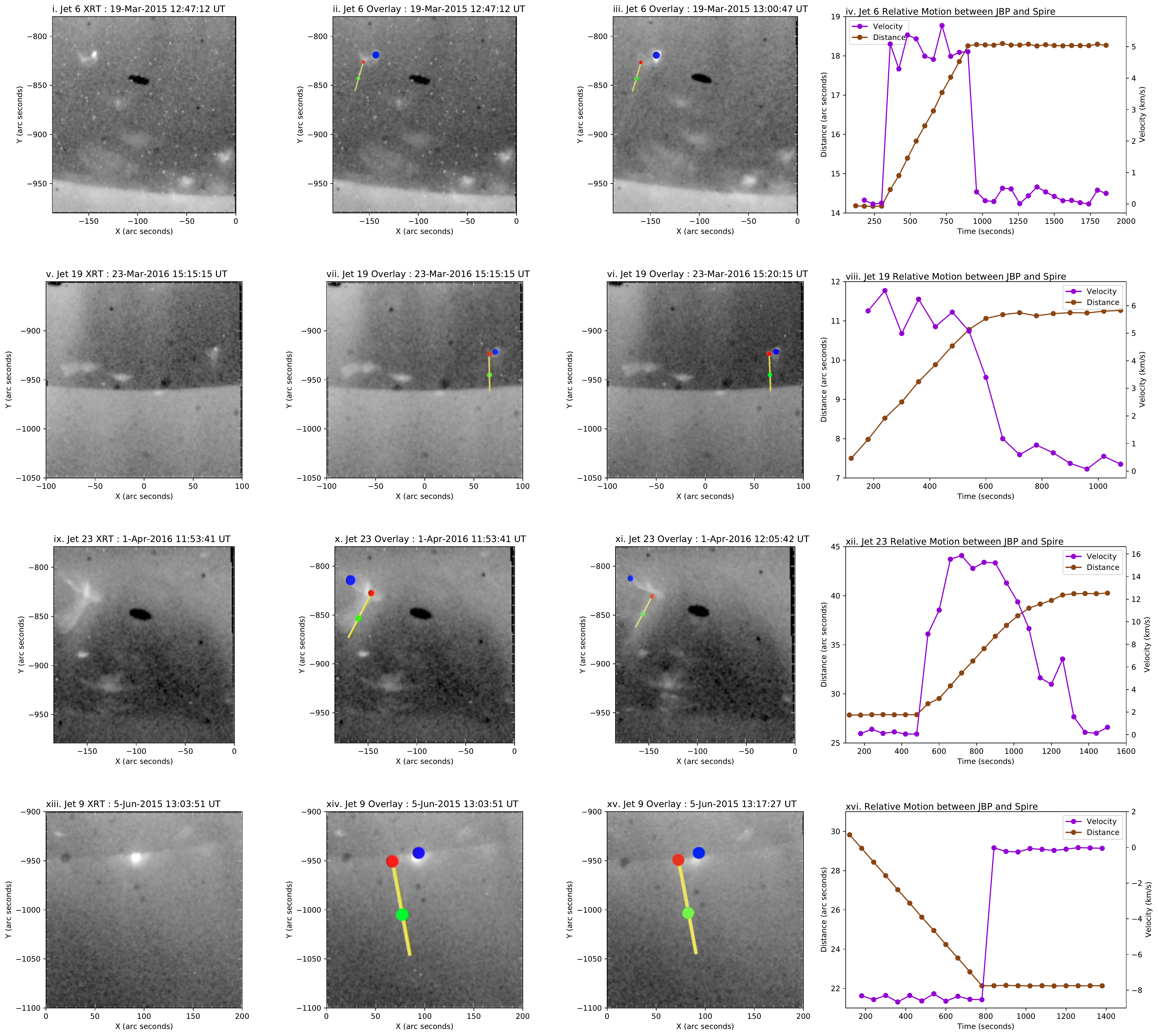}
\caption{Same as in Fig.~5, but for jets 6, 19, 23 and 9 of Table~1.  The first three jets 
are typical of 22 of 23 jets of Table~1, in that all of them show either a virtually 
stationary spire (four of the 23 cases with velocity $\ltsim$1\,\kms), or 
the spires move away from the JBP with time (18 of the 23 cases).  Jet~9 is the lone 
example among the 23 that
shows spire movement {\it toward} the JBP with time (that is, it has decreasing distance from the 
JBP with time).  As is apparent in panels (xiii)---(xv) however,
the jet occurred just at (and likely just beyond) the limb.  Thus there is a strong possibility that the motion of
the spire toward the JBP is only apparent, resulting from lateral swinging of the drifting spire
about the JBP, seen in projection along the Earth-Sun line-of-sight, rather
than actual physical movement of the spire toward the JBP\@.  In the right-hand panels, the respective 
times for the earliest plotted point for jets 6, 19, 23, and 9 are 12:40:12~UT, 15:14:15~UT, 11:40:05~UT, and 13:01:51~UT, where the dates are given in the left-side frames and in Table~1.}
\label{jet_panels}
\centering
\end{figure}
\clearpage

\begin{figure}[]
\includegraphics[width=0.5\textwidth] {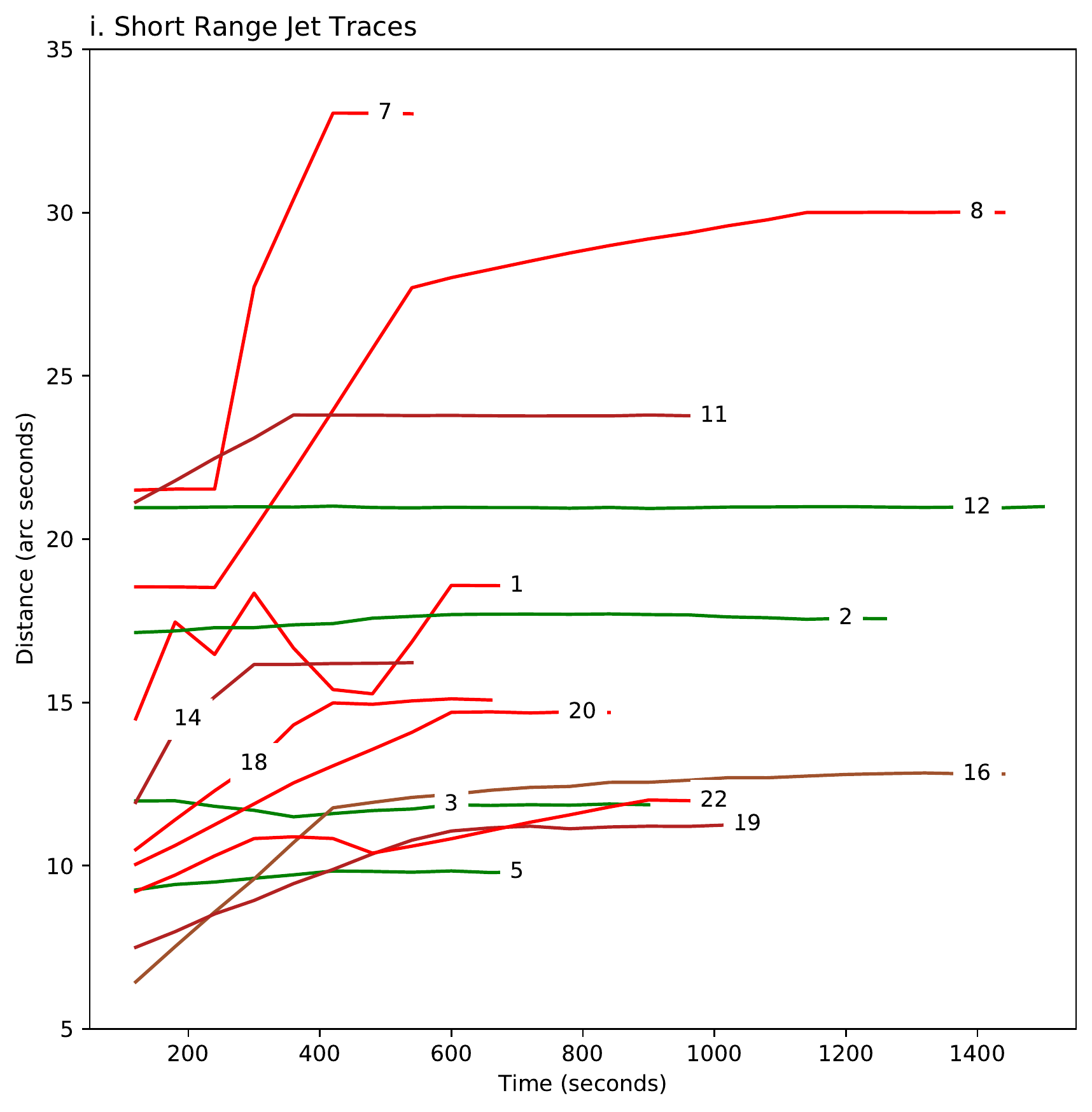}
\includegraphics[width=0.5\textwidth] {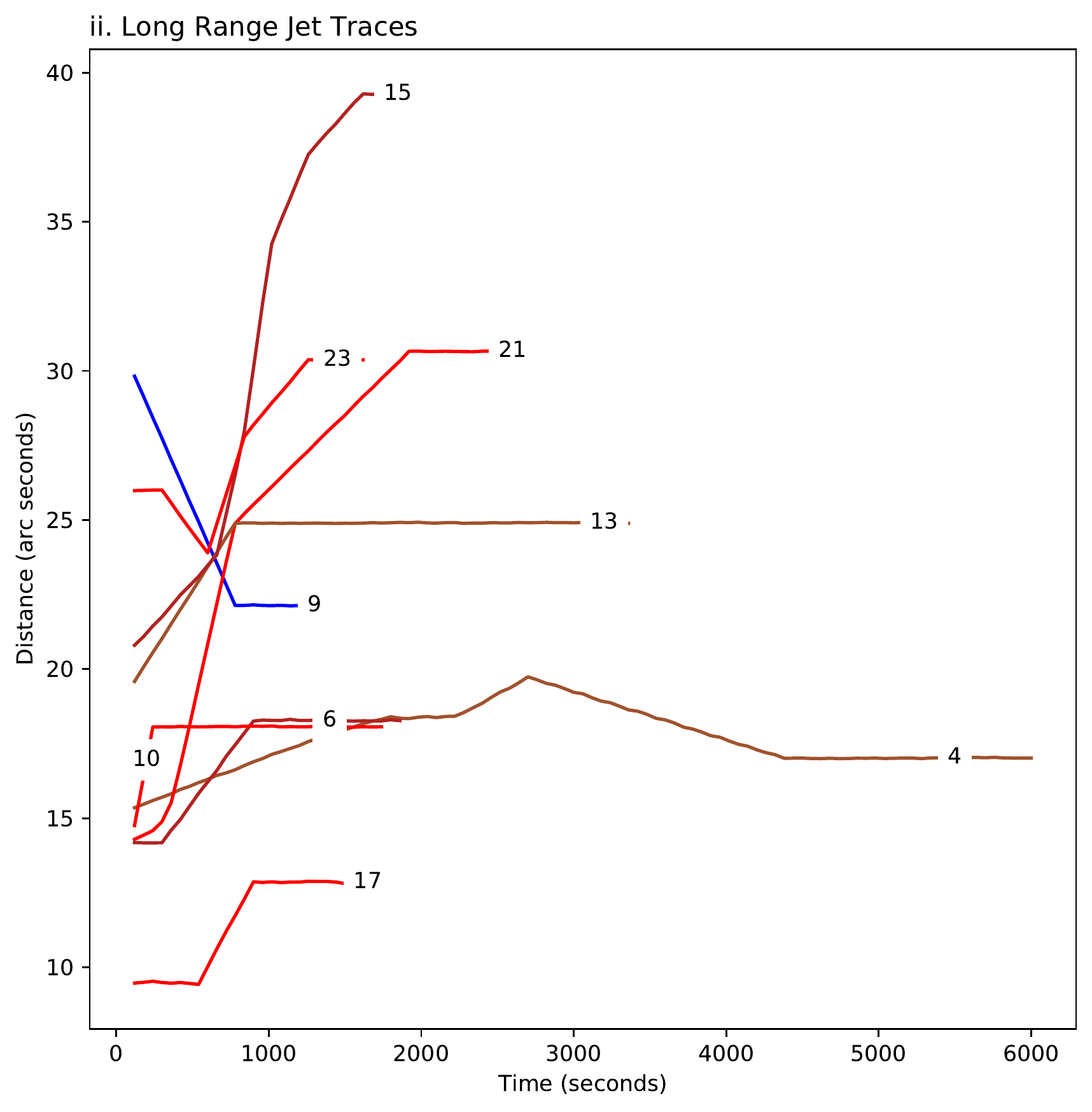}
\caption{Tracks of the lateral distance from the base of the spire to the jet bright point (JBP)
as observed in \hinode/XRT soft X-ray images, as in the brown lines in the right-most panels of Figs.~\ref{al_215_panel} and~\ref{jet_panels}. 
Start times on the horizontal axis are the times given for the jet start in XRT listed in Table~1
for the corresponding jet.
Red, brown, and purple tracks show where the spire-JBP distance increases with time, green tracks represent spires that remain stationary with time (speed $\ltsim$1\,\kms), and the blue line is for
the lone spire for which that distance decreases with time.  All spires, bar one, are either stationary or undergo an increase in the relative 
distance between the JBP and the spire base, consistent with the minifilament-eruption scenario as described
in \S\ref{sec-spires}.  This one 
exception could be a consequence of line-of-sight projection effects of the spire's drift viewed at the solar limb.} 
\label{jet_distances}

\centering
\end{figure}
\clearpage

\begin{figure}[H]
\hspace*{2.2cm}\includegraphics[angle=0,scale=0.80]{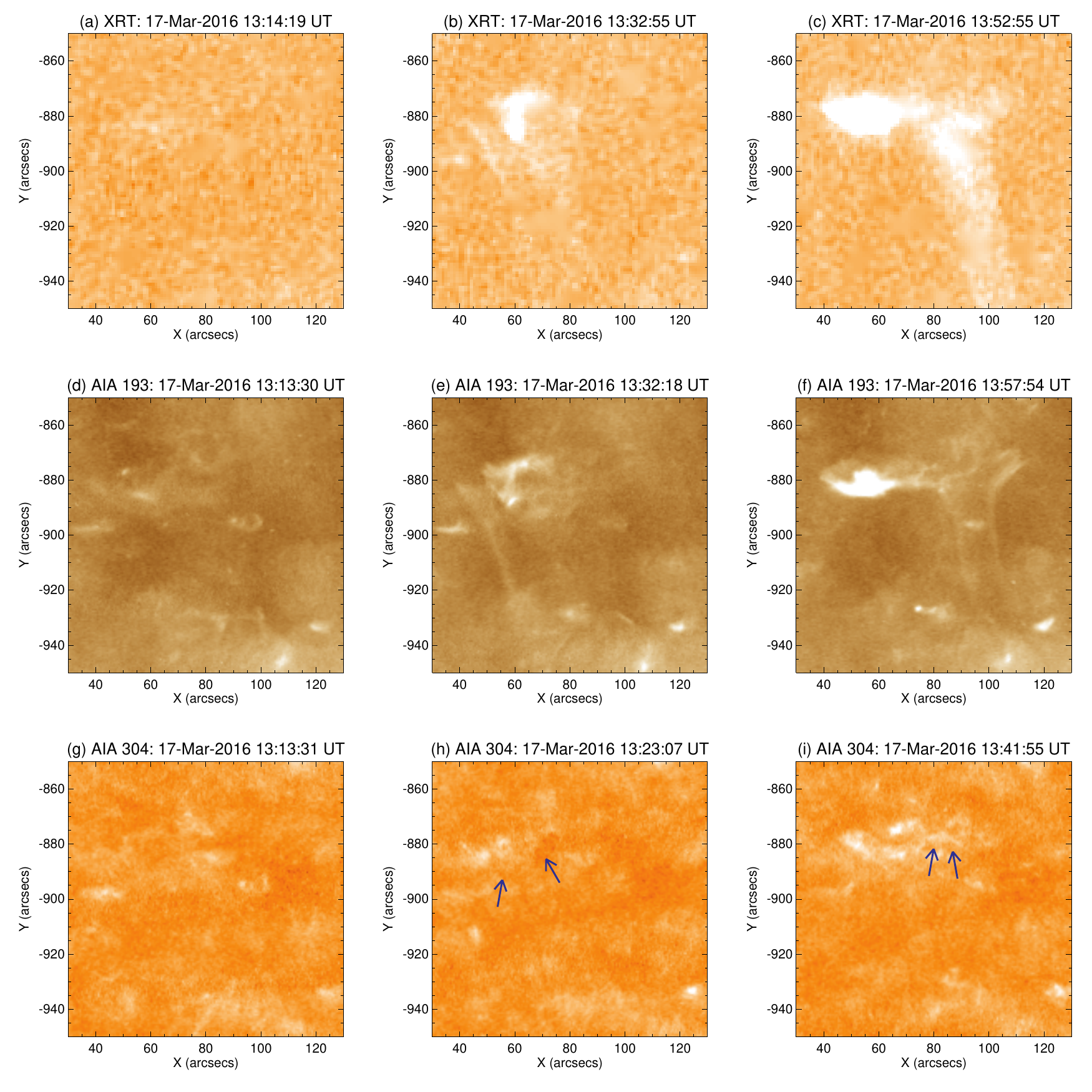}\vspace{1.0cm}
\caption{Sympathetic jets. Jet 15 of Table~1, which consisted of a series of two jets closely linked in space and time.  Panels~(a)---(c)
are from XRT, panels (d)---(f) are from AIA~193\,\AA, and panels (g)---(i) are from AIA 304\,\AA\@.  The first vertical column of
panels shows the situation before either jet starts, the middle column shows the first jet, and the third column shows the
second jet.  The accompanying videos, in movie\_fig8, show the progression from the first 
jet to the second jet in the respective wavelengths.  These are running-sum movies, displayed as described for
the Fig.~3 movies with cadences of 60\,s and 24\,s for XRT and AIA respectively.}
\label{sympathetic_jets}
\centering
\end{figure}
\clearpage

\FloatBarrier


\begin{center} 
\large{Animation Captions}
\end{center}

\noindent
Caption for Video ``movie\_fig3.mp4."  XRT and AIA movies of Jet~1 of Table~1, a standard jet.  The left-panel animation corresponds to Figures~3a, 3b, and 3c, showing 
a \hinode/XRT Al/poly movie of the jet.  This video is constructed by performing a running sum of every
two consecutive images.  This XRT video is presented as a full-cadence (60\,s) running-sum movie.  The center-panel animation corresponds to Figures~3d, 3e, and 3f showing the jet observed by the \sdo\ AIA~211\,\AA\ channel, and the right-panel animation corresponds to Figures~3g, 3h, and 3i, showing 
the jet observed by the \sdo\ AIA 304\,\AA\ channel.  These AIA videos are also constructed by performing a running sum of every
two consecutive images.  For these two AIA movies, however, we
show every-other frame of the running sum, resulting in a cadence of 24\,s, which is still adequate 
to show details of the jet evolution. The movies cover 2014 August 13 over approximately 09:27---09:58\,UT, and the three panels are approximately 
in sync (the exact times are given at the top of each panel). The entire
movie runs in $\sim$6~s.

\vspace{0.5cm}

\noindent
Caption for Video ``movie\_fig4.mp4."  XRT and AIA movies of Jet~2 of Table~1, a blowout jet.  The left-panel animation corresponds to Figures~4a, 4b, and 4c, showing 
a \hinode/XRT Al/poly movie of the jet.  This video is constructed by performing a running sum of every
two consecutive images.  This XRT is presented as a full-cadence (60\,s) running-sum movie.  The center-panel animation corresponds to Figures~4d, 4e, and 4f showing the jet observed by the \sdo\ AIA~193\,\AA\ channel, and the right-panel animation corresponds to Figures~4g, 4h, and 4i, showing 
the jet observed by the \sdo\ AIA 304\,\AA\ channel.  These AIA videos are also constructed by performing a running sum of every
two consecutive images.  For these two AIA movies, however, we
show every-other frame of the running sum, resulting in a cadence of 24\,s, which is still adequate 
to show details of the jet evolution. We use a non-standard colour table for
the AIA~304\,\AA\ animation, as we found our selected colours to show the 
erupting minifilaments more clearly for these cases than when rendered with the standard 
colour table.  The movies cover 2015 February 27 over approximately 12:14---13:13\,UT, and the three panels are approximately in sync (the exact times are given at the top of each panel). The entire
movie runs in $\sim$6~s.

\vspace{0.5cm}

\noindent
Caption for Video ``movie\_fig8.mp4."  XRT and AIA movies of Jet~15 of Table~1, which, although listed as one jet in the table, actually consisted of a series of two jets closely 
linked in space and time (sympathetic jets).  The left-panel animation corresponds to Figures~8a, 8b, and 8c, showing 
a \hinode/XRT Al/poly movie.  This video is constructed by performing a running sum of every
two consecutive images.  This XRT is presented as a full-cadence (60\,s) running-sum movie.  The center-panel animation corresponds to Figures~8d, 8e, and 8f showing the jet observed by the \sdo\ AIA~193\,\AA\ channel, and the right-panel animation corresponds to Figures~8g, 8h, and 8i, showing 
the jet observed by the \sdo\ AIA 304\,\AA\ channel.  These AIA videos are also constructed by performing a running sum of every
two consecutive images.  For these two AIA movies, however, we
show every-other frame of the running sum, resulting in a cadence of 24\,s, which is still adequate 
to show details of the jet evolution. We use a non-standard colour table for
the AIA~304\,\AA\ animation, as we found our selected colours to show the 
erupting minifilaments more clearly for these cases than when rendered with the standard 
colour table.  The movies cover 2016 March 17, over approximately 12:50---14:25\,UT, and the three panels are approximately in sync (the exact times are given at the top of each panel). The entire
movie runs in $\sim$9~s.

\vspace{0.5cm}

\end{document}